# Risk-Driven Compliant Access Controls for Clouds

*Hanene Boussi Rahmouni, Kamran Munir, Mohammed Odeh, Richard McClatchey*

Department of computer science and creative technology, FET, UWE, Bristol, UK

**Abstract**

There is widespread agreement that Cloud computing has proven cost cutting and agility benefits. However, security and regulatory compliance issues are continuing to challenge the wide acceptance of such technology both from social and commercial stakeholders. An important factor behind this is the fact that Clouds, and in particular public Clouds, are usually deployed and used within broad geographical or even international domains. This implies that the exchange of private and other protected data within the Cloud environment would be governed by multiple jurisdictions. These jurisdictions have a great degree of harmonisation, however, they present possible conflicts that are difficult to negotiate at run time. So far, important efforts have been taken in order to deal with regulatory compliance management for large distributed systems. However, measurable solutions are required for the context of Cloud. In this position paper, we propose an approach that starts with a conceptual model of explicit regulatory requirements for exchanging private data on a multijurisdictional environment and build on it in order to define metrics for non-compliance or risks to compliance. These metrics will be integrated within usual data access-control policies and will be checked at policy analysis time before a decision to allow/deny the data access is made.

## 1. Introduction

Globalisation has changed the way societies and organisations work. As a consequence, technologies have had to evolve in order to be able to cope with new organisational needs. New concepts such as Cloud computing have emerged as approaches to mitigating the needs and facilitate those changes. However, the governance of such systems has to evolve for a multijurisdictional context. The complexity of such governance context makes it challenging to abstract and enforce applicable high level regulatory requirements as operational policies. On the plus side, if these challenges are tackled measurable compliance coverage on the Cloud would be possible. In this paper, we present an approach to bridging the gap between international regulatory frameworks on systems and data security and operational level policies for accessing computer resources through Cloud services.

This approach is based on the semantic modelling of explicit legal requirements, which will be extracted from international legislation on systems and data security. This includes the Sarbanes-Oxley (SOX) initiative in the United States and the Data Protection directive of the European Union. In this research, the aim is to build a generic model to enable the specification of these requirements in a standard way. In addition, the model should reflect both similarities and conflicts between the different legislations forming part of the Cloud's governance framework. We further extend our model to capture prioritised non-compliance risks. This is done through the use of semantic rules that are complementary to our model. We finally integrate these risks as access control conditions. The satisfaction of these conditions would lead to denial of access to protected resources.

The work on this paper builds on our previous efforts of an ontology-based approach to privacy compliance for the sharing of Health data in Europe. When completed, this work should add useful contributions to the area of regulatory compliance on Clouds domains. This includes: (1) the specification and formalisation of requirements extracted from diverse jurisdictions through the use of highly expressive semantic languages such as SWRL, RIF and LKIF; and (2) through the use of these formalisms, enabling the detection of situations of data disclosures or access that might present risks to regulatory compliance. This would allow the prevention of non-compliance at access control run time.

This paper is structured as follows. After presenting an introduction in Section 1, Section 2 describes the proposed approach. In Section 3, we present the methodology which we are following to undertake this research. Section 4 covers the related work; and finally, the conclusions and future work are presented in Section 5.

## 2. Description of Proposed Research

Cloud computing allows efficient cost cutting through the use of custom-tailored, on-demand IT solutions. However, Cloud providers are still failing to guarantee measurable handling of many non-functional requirements, such as security and regulatory compliance [4]. Although they are newly emerging paradigms, Clouds are very similar in many aspects to other distributed computing environments. In particular, Clouds are similar to large scale systems that are based on virtualised technologies such as grid systems [24].

These systems often fall short of providing measurable proof of compliance. The presence of such shortcomings may lead to decreased social trust and acceptance and therefore low business engagement with the Cloud business case [21]. In this research, we aim to specify and to enforce access control policies that allow the identification of compliance vulnerabilities at policy evaluation time. In our context, situations of compliance vulnerabilities will be specified through extending a model of traceable requirements extracted from high level policies. This latter is to be specified at an early stage of the research. Finally, the captured risks will be integrated as access control conditions.

We aim to achieve these research objectives by following a phased but iterative approach shown as Figure 1. In the first phase, we shall extract enforceable requirements from different security and trust legislation that form part of the legal framework of Cloud computing, including SOX [23] and privacy and data protection legislation. Due to the diversity and scope of this legislation, this research starts with a deep analysis of the legal requirements for sharing protected resources on the Cloud. Throughout this analysis, similarities as well as possible space for conflict between different legislation will be identified. We also consider for scalability concerns, identifying a selection of legal requirements with a high priority for enforcement.

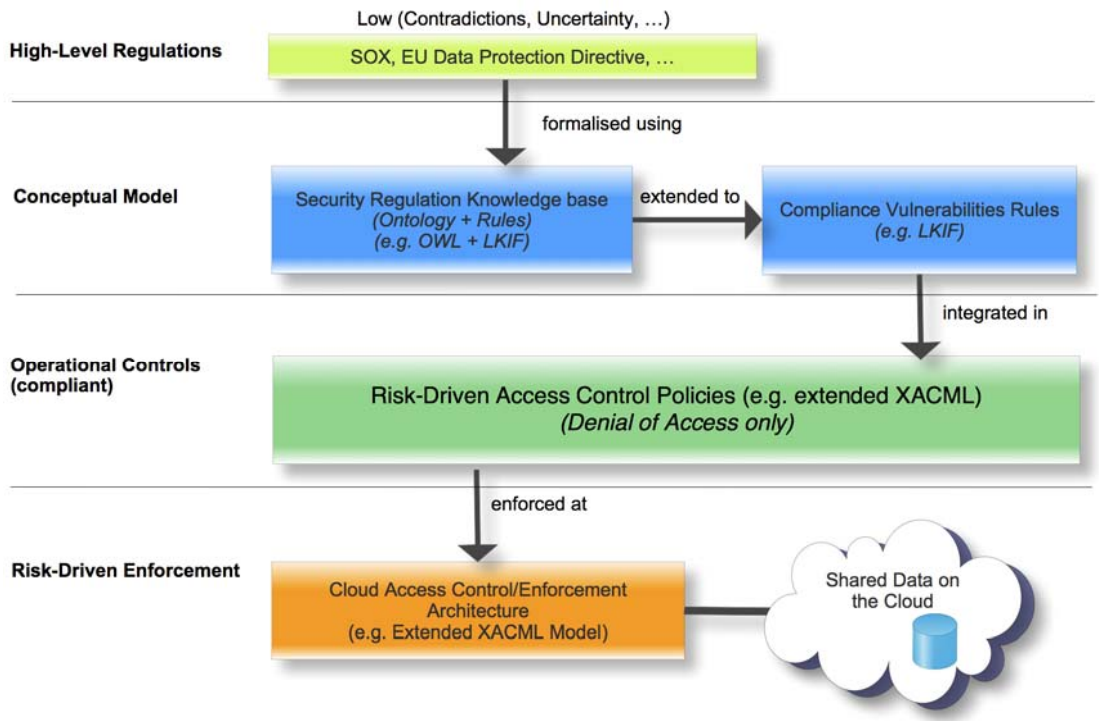

**Figure 1**: A Proposed Roadmap to Enforce Compliant Access Control (Denial of Access)

In the second phase, we conceptually model the legal requirements identified in the previous phase. In addition, we specify and model situations of non-compliance with these requirements. The modelling language we are using for these tasks should be expressive enough in order to handle uncertainty and contradictions of legislation. The complexity of such issues scales when working on multijurisdictional and/or multicultural domains of which we believe internationalised CClouds are an example. It is therefore necessary to look at a semantic language that is the most expressive and the most complete among the other languages constituting the semantic web stack [29]. One possible choice could be for the Web Ontology Language (OWL) and the Semantic Web rule Language (SWRL). However, in general (independently from the chosen rule language) the rules would be specified in the form of production rules, which capture information about the generic case of data disclosure as rule antecedent and dictate either a single legal requirement or a conjunction of requirements in their consequence. The following is an example of a privacy requirements rule indicating the necessity of obtaining patient consent when disclosing private data coming originally from France:

    *dataSharing(?x)*
∧ *concerning(?x, ?data)*
∧ *belongsto(?data, France)*
∧ *isPrivate(?d)*
→ *hasConsentNecessityRequirement(?x,*
    *NecessaryConsent)*

In addition to requirements specification rules we also define rules that derive metrics for detecting risks to compliance with a legal requirement. For this reason, we need first to define metrics for testing for the satisfaction of each requirement. For example, we define a metric: *PatientConsent(X, obtainedConsent)*; presented in the form of an OWL axiom that checks for the satisfaction of the consent necessity requirements for a specific data disclosure *X*. In the cases where the requirement of necessary consent is not satisfied this means that the patient consent was not obtained prior to the data disclosure or the consent was obtained to operate for a specific period of time but has expired. Hence, one of the following axioms will hold true *PatientConsent(X, NoConsent) or PatientConsent(X, ExpiredConsent)*. The applicable risk derivation rules for this example are presented as follows:

*dataSharing(?x)*
∧ *hasConsentNecessityRequirement(?x, NecessaryConsent)*
→ *hasComplianceRisk(?x, NoConsent)*

OR

*dataSharing(?x)*
∧ *hasConsentNecessityRequirement(?x, NecessaryConsent)*
→ *hasComplianceRisk(?x, ExpiredConsent)*

The work we have done previously in (Rahmouni, 2011) attempted to model and formalise EU legislation on data protection through an OWL plus SWRL knowledgebase, while addressing some of the ambiguities and contradictions characterising it. This has allowed reasoning about policies from different European Member States (Rahmouni, 2009). However, despite its high expressiveness capability, SWRL has some limitations in fully handling the non-monotonicity of legislation. In other words, the absence, negation and other non-monotonic operations deprives SWRL from natively reasoning on conflicting rules.

In order to overcome this efficiency we had to programmatically set a priority for rule execution by borrowing the "salience" keyword which is a reserved keyword of the rule engine that was used to execute the rules. In this research, we intend to explore an extension to SWRL or a change to a different semantic rule language such as RIF or LKIF. The work in (Gordon et al., 2009 and Palmirani et al., 2009) provides a comparison of the ability of different semantic rule languages for regulation modelling. Also the work in (Estrella, 2010) proposes LKIF, an extension to SWRL, as a better approach to dealing with non-monotonicity issues of legislation. This is because the LKIF rule language includes a built-in predicate which is enforced over rules and allows reasoning about rule priority for execution where conflicting rules are found.

In the third phase, the semantically captured compliance risks could be mapped to metrics forming part of the access control conditions that are usually specified in a standard access control language such as XACML. In [20] and [21], we have shown how we model high level policies interpreted from European and national data protection law as privacy compliant access control policies. The use of semantic web technologies such as OWL and SWRL to allow integration of privacy constraints extracted from law as policy constraints, such as the requirements of consent and other safeguards of patient rights. In this previous work, we have checked for compliance at policy evaluation time by considering the positive authorisation mode. We believe that the process of checking for non-compliance, as suggested in this paper, would be more practical and efficient than the process of checking for compliance. For example, consider a specified sharing *X* of some protected resources on the Cloud. We suppose that *X* has a set of legal requirements with which a user must comply. In order to decide whether *X* is compliant, we need to check that each of the applicable requirements was satisfied. If so *X* would be permitted. In contrast, fewer checking operations are needed in order to measure non-compliance. This is because it is sufficient to know that at least one requirement from the set of requirements applicable to *X* was not satisfied, in order to infer non-compliance.

At a later stage, the mapping of compliance risks to access control metrics would facilitate the integration of our approach within existing access control models, which manage access to protected data on CCloud infrastructures. However, these legacy models might require extensions in order to be able to process complex and dynamic constraints specified in our data access and usage polices. This is handled in the next phase.

Finally in a fourth phase, we will design a prototype architecture for the enforcement of the suggested access control. In addition, we will implement the prototype and test it with regards to a specific case study (e.g. business enterprises, healthcare, etc.). We believe this approach would provide a preventative method for compliance management. Moreover, it provides mechanisms to allow enterprises working with Cloud infrastructures to adopt a measurable way of enforcing high level legislation within their operational processes. The abilities to detect vulnerable disclosures and to prevent them happening would also increase the level of social trust and acceptance of the Cloud paradigm. Furthermore, a tailored retrospective mechanism for auditing compliance at the time of access or usage control will also be considered in the future.

## 3. Research Methodology

For this research, we aim to follow the structured framework of Legal Knowledge Engineering and Enforcement '[REF?]. This framework will be driven by an ontology of predefined data disclosure models along with an ontology of legal privacy and security requirements. In addition, our research framework includes three more layers that lead to the final stage of enforcing risk-aware access control policies in the Cloud operational processes. In total, our framework contains five layers as shown in Figure 2. In the following sections we present these layers in more detail.

Most layers of this framework synchronise with the different phases of TUREP [15], which is a software engineering process that is very close to a universal and integrated requirement engineering framework. The TUREP model was proven successful in paving the way for a universal requirements engineering process through the involvement of its key characteristics as adhering to the engineering dimension. In particular, TUREP is in line with other research projects that have attempted to bridge the gap between business processes and system/data models and requirements specifications. These include BPMOntoSOA [32] (a semantic-based approach to generating service oriented requirements specifications from business process models), OntoREM [13, 14] (an ontology or knowledge driven requirements engineering methodology applied in the aerospace sector), OntoRAT [2] (an ontology driven requirements analysis tool) and generating conceptual data models from knowledge models such as OWL-DL models [7]. Therefore, these research projects are considered as part of the building blocks to developing an integrated but generalised framework with the TUREP model at the centre in attempting to evolve a universal integrated framework for requirements engineering.

However, in order to satisfy the research problem discussed in this paper our framework includes an additional layer where we look at mapping legal requirements onto metrics representing risks for non-compliance. These are also called compliance vulnerabilities.

It is also worth mentioning before proceeding into the detailed description of the framework's layers that due to the diversity and complexity of the different legislation (that forms part of the governance of disclosure of protected data across international borders) we consider restricting our focus to privacy legislation only, in particular, legislation governing the exchange of health data. This includes for example, the European Directive on Data protection and some US legislation such as the Health Insurance Portability and Accountability Act (HIPAA). The following sections present more details on the framework layers (as shown in Figure 2).

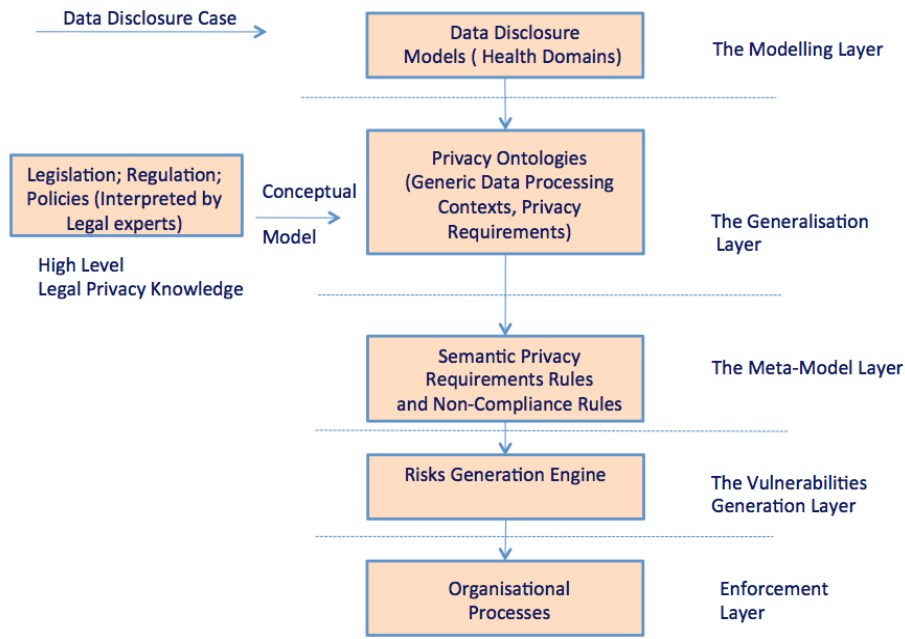

**Figure 2**: An Ontology-Driven Framework for Legal Requirements and Risks Engineering for the Exchange of Private Data on a Cloud Environment

### 3.1.1. The Modelling Layer

The modelling layer focuses on the design and production of a semantic model or an ontology of specified cases of data disclosures that take place within an international Cloud environment. In particular, we semantically represent cases of disclosures of data revealing information about a person's health. Moreover, in this phase we also gather generic requirements described in text law. These requirements are then analysed according to their relevance to our domain (i.e. the health domain and the relevance of the data type to cases of data disclosures we are interested in). Furthermore, due to the need to narrow down the scope to the time consuming nature of semantic modelling, the gathered requirements will be prioritised according to their importance to the task of effective data protection and privacy preservation. For example, the requirements of obtaining patient consent prior to disclosing their data seems to have higher importance compared to data anonymisation, which is important but not always required. After prioritisation, these requirements will be captured in an ontology or knowledge base. Both ontologies (i.e. the data disclosure context ontology and the privacy requirements ontology) set up the context for the domain for which we later formally specify applicable and fine grained legal requirements.

This phase is in line with the initial phase of the TUREP model in which the requirements should be gathered, elicited and analysed. The next layer of the framework focuses on the formal specification of legal requirements and risks to compliance.

### 3.1.2. The Generalisation Layer

In this layer, we effectuate a set of tasks that focus on the generalisation of the context of data sharing or the disclosure which were specified in the previous layer. Each specific case of data disclosure will be mapped to a generic data processing context that is specified in text law. This is in order to conform to the nature of legislation that is generic in the way it is written, but has to be implemented or enforced in a case based manner. In this regard, we generalise our specialised cases of data disclosure in order to be able to match each of them onto a set of applicable legal requirements that has to be complied with before the disclosure is permitted.

### 3.1.3. The Meta-Model Layer

This layer illustrates a phase in which generic templates for requirements specification are formulated. Since we are dealing with legal requirements which are by nature specified, in text law, in the form of rules, it is more logical to formulate these requirements through the use of formal or logical rules, notably production rules. These rules capture information about the generic case of data disclosure and would dictate either a single requirement or a conjunction of requirements in their consequence. These generic rules will be instantiated, at a later stage, in order to specify the requirements applicable for specialised cases of data disclosure.

In addition to requirements specification we also specify risks to compliance with these requirements in a way that allows, for each production rule specifying a requirement, associating it with a set of other rules defining risks to compliance with that specific requirement. The requirements specification and risk derivation rules will be used in the next phase of the framework in order to automatically generate applicable risk detection metrics. These metrics will serve to identify attempts to non-compliance at data access runtime.

### 3.1.4. The Compliance Vulnerabilities Generation Layer

In this layer, the rules produced as an output of the previous phase will be executed through adequate rule engines in order to derive applicable legal requirements. For example, in the context of disclosing a patient's chest MRI, where the MRI was originally collected from a French hospital, the applicable requirements along with risks to complying with these requirements will be generated (as described in Figure 3).

| Data Disclosure Subject | Legal requirements | | Legal Risks | |
|---|---|---|---|---|
| Type of Data to be disclosed: patient's *chest MRI* | Country: France | | | |
| | **Consent Necessity** | | **Consent Necessity** | |
| | Necessary-Consent | | No-Consent or Expired-Consent | |
| | **Consent Specificity** | | **Consent Specificity** | |
| | Specific Consent | | Broad Consent | |
| | **Consent Explicitness** | | **Consent Explicitness** | |
| | Explicit-Consent | | Implicit-Consent | |

**Figure 3**: Example of legal requirements and risk derived for a given disclosure of a patient's MRI.

### 3.1.5. Enforcement Layer

The fifth and final layer of our research framework looks at the enforcement of compliance requirements through the integration of risk metrics within the specification of access control policies governing the operational processes of the Cloud's platforms. These metrics will be included in the conditions, which normally form part of formal policies specification. The conditions describing the satisfaction of a risky situation or compliance vulnerability will be checked at run time i.e. when a policy decision is being made. If at least one risk metric is validated during this check then the applicable decision will be a denial of access and/or disclosure of the protected data.

## 4. Related Work

Since there is only limited work on compliance management in Clouds, in this section we focus on related areas such as compliance management in general i.e. in the software development life cycle and at policy enforcement time. Moreover, we also cover the state-of-the-art in formal policies and regulation specification and enforcement languages.

### 4.1. Compliance Management Approaches

Most of the existing work in the area of compliance management deals with a particular problem of compliance auditing. This includes for example, Hippocratic databases [1] and some semantic web-based approaches such as [31] and [9]. These approaches however are not sufficient for our study. We believe that in order to achieve effective compliance management we need not only to rely on retrospective approaches, but complementary proactive approaches are also needed in order to show adoption of good practice and more assurance for compliance. Examples of these approaches are described in [22], [20] and [11]. The work in [22] and [20] had particular emphasis in European privacy legislation, whereas Gowadia [11] looked at integrating requirements interpreted from data-sharing agreements. We believe that relying on secondary legislations such as contracts, protocols and agreement does not guarantee completeness of compliance when compared to primary text law. The research presented in [22] does not check for compliance with explicit fine grained legal requirements extracted from legislation, rather only generic requirements interpreted from high level legal principles have been considered.

In particular, our work presented earlier in [21] attempts to model and formalise EU Legislation through an OWL plus SWRL knowledgebase. The approach however had some limitations linked to the inability of SWRL to fully handle the non-monotonicity of legislation. The reason for these limitations was explained in section 2 of this paper. In this current research, we intend to explore an extension to SWRL or a change to a different semantic rule language such as RIF or LKIF in order to deal with these limitations.

In [20] and [21], we have shown how we model high level policies interpreted from European and national data protection law as privacy compliant access control policies. The use of semantic web technologies such as OWL and SWRL allowed the integration of privacy constraints extracted from law, such as requirements of consent and other safeguards of patient rights, as policy constraints. Also, this solution has permitted the checking for requirements satisfaction at run time and makes an access control decision according to the finding. We believe it would be more practical and less time demanding if we manage to detect, at run time, situations of non-compliance and prevent them happening. That is the object of this work.

### 4.2. Formal Policy and Regulation Enforcement Languages

High level legal rules need to be interpreted at an operational level so that specific decisions induced by events can be taken. For this purpose, we need to investigate the expressiveness and ability of some promising policy languages that have been advantageous in the mapping of requirements from privacy laws and also guidelines into operational controls. After going through some data protection legislation, we noticed that the legislation mainly dictates requirements and expectations that need to be fulfilled by a data controller throughout the whole lifecycle of the management of the personal data in question. These requirements could be classified into two categories (1) obligations related to the access and usage of personal data based on the data subject's consent and/or complying with pre-identified usage purposes and (2) obligations to be fulfilled on the collected data such as data retention, deletion and correction, notification to the data subject, secondary use of data and data transfer to third parties.

A considerable amount of work has been done in the area of security policy languages. However, many of these approaches have mainly captured the access control aspects of these obligations and do not focus on the duties and the expectations on how personal data has to be handled. Some approaches have shown awareness of the need to deal with other aspects of the obligations than just access control obligations but they usually stop at the specification of these controls or tried to enforce them only at local domains where the data resides. This gives less control and traceability on how data will be handled after the access is permitted and data is transferred to an external domain. In the following sections, a review of different categories of formal policy languages is presented that have been used to enforce security and privacy controls.

#### 4.2.1. Semantic-based Languages

A semantic-based description of policies through an ontology allows easy access to the policies' information and facilitates capturing the relations between policies, entities and other policies. It is also possible to query an ontology in order to get useful information. This is rather different from traditional languages that define only predefined queries in order to access information about policies. However, most of the languages that have been designed so far rely on a pure OWL syntax. These policies present some limitations in terms of expressiveness, which is in turn inherited from known limitations. Moreover, as in all ontology-based systems, keeping up with version changes to the ontologies as well as the integration and interoperability between different ontologies in the same domain continues to be an open research area.

Rei [12] is a policy framework that allows specification, analysis and reasoning about declarative policies. Rei adopts a rule-based approach to specifying semantic policies. Rules are expressed as OWL properties of the policy and are classified as rights, prohibitions, obligations and dispensations. Though represented in OWL-Lite, Rei still allows the definition of variables that are used as placeholders, as in Prolog [26]. The definition of constraints follows the typical pattern of rule-based programming languages, like Prolog, i.e. defining a variable and the required value of that variable for the constraint to be satisfied. In this way, Rei overcomes one of the major limitations of the OWL language and more generally of description logics, i.e. the inability to define variables. On the other hand, the choice of expressing Rei rules similarly to declarative logic programs prevents it from exploiting the full potential of the OWL language. OWL inference is essentially considered as an oracle, i.e. the Rei policy engine treats inferences from OWL axioms as a virtual fact base. Hence, Rei rules cannot be exploited in the reasoning process that infers new conclusions from the existing OWL ontologies, which means that the Rei engine is able to reason about domain-specific knowledge, but not about policy specification. Therefore, in order to classify policies, the variables in the rules need to be instantiated, which reduces it to a constraint satisfiability problem.

The KAoS [25] framework through its policy services allows the specification, management, conflict resolution and enforcement of policies within agent domains. KAoS adopts an OWL-DL ontology-based approach to describing and specifying policies and context conditions. The ontological modelling of the policies helps reasoning about both domain and policy specification and facilitates the detection of policy conflicts at policy definition time. However, a pure OWL approach encounters some difficulties with regard to the definition of some kinds of policies. Also the gap between the specification and the actual implementation of such policies cannot be coped with automatically [27]. The lack of facilities to describe variables makes it difficult to define policies that contain parametric constraints, which are assigned a value only at deployment or runtime. For this reason, KAoS developers have introduced role-value maps as OWL extensions, which allow KAoS to specify constraints between properties' values expressed in OWL terms and to define policy sets i.e. groups of policies that share a common definition but can be singularly instantiated with different parameters. The proposed extensions effectively add sufficient expressive flexibility to KAoS to represent the policies discussed in this paper. However, non-experienced users may still have difficulties in writing and understanding these policies without the help of the KPAT [25] graphical user interface.

### 4.2.2. XML-based Privacy Policy Languages

Some XML-based policy languages were formalised originally for the purpose of specifying operational controls to the access of protected data in general, but they evolved to formally represent some policies of legal aspects and, in particular, legal privacy and data protection policies.

The work on the Platform for Privacy Preferences Project by W3C has lead to the production of the P3P [16] specification. P3P enables users to express privacy practices of their websites in a standard format that can be retrieved automatically by user agents. P3P user agents allow users to be informed of site practices (in both machine and human-readable formats) and to automate decision-making based on these practices when appropriate [5]. However, there are some shortcomings associated with the enforceability of P3P policies such as (1) predefined types of purposes for allowing access to personal data make it difficult for organizations to define their own purposes (2) unavailability of read, write, delete, append actions (3) P3P does not use or specify any post processing obligations such as notification to data subject, or retention of data for a specified period and (4) P3P has limited conditions of data processing i.e. simple opt in/ opt out conditions [3].

The Enterprise Privacy Authorization Language (EPAL) [18] was designed to enable the translation of privacy policies into an XML based computer language. The resulting coded translation of human policy into Information Technology policy allows a complex description of the internal data handling practices needed for enforcing the privacy policy. Similarly, XACML [30] is an XML specification and syntax for expressing policies controlling the access of information through the internet. XACML provides enterprises with a flexible and structured way of managing access to resources. XACML is based on a subject-target-action-condition policy syntax applied to XML documents that can update individual document elements. Like other policies languages that are based on XML, XACML lacks the required semantics to allow for semantic heterogeneity and interoperability especially when managing data access within environments that involve multiple organisations.

### 4.3. Other Formal Policy Languages and Techniques

Other types of formal policy language that have recently emerged and that are worth mentioning aredeclarative languages that are written on a programming language syntax and are easier to integrate or embed in code snippets while developing applications or services that are meant to handle or manage protected data. One of the most famous examples of these languages is Ponder [6], which is a declarative object-oriented language for the specification of management and security policies. It provides structuring techniques to serve the complexity of policy administration in large enterprise information systems. It has been recently extended to a more advanced version named Ponder2 [28]. This language offers a wide range of interesting models of concepts such as hierarchical data structure for organizing managed objects, a policy evaluation points (PEPs) scheme, and their conflict resolution strategy. Fundamentally, Ponder2 distinguishes between two policy types that are obligations and authorisations. Usually the obligation policies define the actions that policy subjects must perform on target entities when specific relevant events occur. Authorisation policies are usually used to specify operations a subject is authorised to do on target objects. Also, Ponder2 includes some composite policy types that allow the grouping of basic policies into the shape of roles or relationships policies. However, the policy language of Ponder2 does not allow definition of obligation actions that must be executed before or after an authorisation.

## 5. Conclusions

The work presented in this position paper has been based on our previous work [20, 21], which presents an innovative approach to privacy compliance management for the sharing of Health data in Europe. In this work, we consider the problem of privacy compliance from a software engineering point of view where compliance is considered as a Non-Functional Requirement for distributed systems deployed at the international level e.g. International Clouds. As stated previously our methodology conforms to the TUREP [15] software engineering model that has proven successful and effective for structuring requirement engineering processes in different domains.

The extension to our previous work include: (1) extending the geographical scope of the covered legislation from an European to an international level (2) adopting a more advance and expressive formal regulation specification language (3) detecting risks to compliance (also called compliance vulnerabilities) through the use of well defined metrics at access control runtime and preventing them from happening and finally (4) testing and validation of the proposed solution on a Cloud environment. Our future work will focus on putting our methodology into practice. For testing and validation purposes, a case study with rich scenarios of private data sharing in healthcare will be carefully selected and applied to the different layers of our requirements engineering framework.